\documentclass[apjl]{emulateapj}%
\usepackage{amsmath}
\usepackage{graphicx}
\newcommand{\uJy}{\mu {\rm Jy}}

\newcommand{\simpl}{{\sc simpl}}

\newcommand{\kerrbbtwo}{{\sc kerrbb2}}

\newcommand{\tbabs}{{\sc tbabs}}

\newcommand{\lowess}{{\sc lowess}}
\newcommand{\smarts}{SMARTS}

\newcommand{\mdot}{{\dot{m}}_X}

\newcommand{\rchinu}{\chi^{2}/\nu}

\newcommand{\rxte}{{\it RXTE~}}

\shorttitle{Accretion Lags in LMC X--3}
\shortauthors{Steiner et al.}

\begin{document}

\title{Modeling the Optical -- X-ray Accretion Lag in LMC X--3: Insights
  Into Black-Hole Accretion Physics}

\author{James F.\ Steiner\altaffilmark{1}\altaffilmark{\dag}, Jeffrey
  E.\ McClintock\altaffilmark{1}, Jerome A.\ Orosz\altaffilmark{2},
  Michelle M.\ Buxton\altaffilmark{3}, Charles D.\ Bailyn\altaffilmark{3},
  Ronald A.\ Remillard\altaffilmark{4}, and Erin Kara\altaffilmark{5}}

\altaffiltext{1}{Harvard-Smithsonian Center for Astrophysics, 60
  Garden Street, Cambridge, MA 02138.}
\altaffiltext{2}{Department of Astronomy, San Diego State University, 5500 Campanile Drive, San Diego, CA 92182-1221.}
\altaffiltext{3}{Astronomy Department, Yale University, P.O. Box 208101, New Haven, CT 06520-8101.}
\altaffiltext{4}{MIT Kavli Institute for Astrophysics and Space
  Research, MIT, 70 Vassar Street, Cambridge, MA 02139.}
\altaffiltext{5}{Department of Astronomy, Cambridge University,
  Madingley Road, Cambridge, CB3 0HA, UK.}
  \altaffiltext{\dag}{Hubble Fellow.}

\email{jsteiner@cfa.harvard.edu}

\begin{abstract}

  The X-ray persistence and characteristically soft spectrum of the
  black hole X-ray binary LMC X--3 make this source a touchstone for
  penetrating studies of accretion physics.  We analyze a rich,
  10-year collection of optical/infrared (OIR) time-series data in
  conjunction with all available contemporaneous X-ray data collected
  by the ASM and PCA detectors aboard the {\it Rossi X-ray Timing
    Explorer}.  A cross-correlation analysis reveals an X-ray lag of
  $\approx 2$ weeks.  Motivated by this result, we develop a model
  that reproduces the complex OIR light curves of LMC X--3.  The model
  is comprised of three components of emission: stellar light;
  accretion luminosity from the outer disk inferred from the
  time-lagged X-ray emission; and light from the X-ray-heated star and
  outer disk.  Using the model, we filter a strong noise component out
  of the ellipsoidal light curves and derive an improved orbital
  period for the system.  Concerning accretion physics, we find that
  the local viscous timescale in the disk increases with the local
  mass accretion rate; this in turn implies that the viscosity
  parameter $\alpha$ decreases with increasing luminosity.  Finally,
  we find that X-ray heating is a strong function of X-ray luminosity
  below $\approx 50$\% of the Eddington limit, while above this limit
  X-ray heating is heavily suppressed.  We ascribe this behavior to
  the strong dependence of the flaring in the disk upon X-ray
  luminosity, concluding that for luminosities above $\approx50$\% of
  Eddington, the star lies fully in the shadow of the disk.

\end{abstract}


\section{Introduction}\label{section:Intro}

LMC X--3 was one of the first extragalactic X-ray sources to be
discovered \citep{Leong_1971}.  It was subsequently identified with a
B main-sequence star \citep{Jones_1974, Johnston_1979}.  With the
discovery that the star is in a 1.7-day binary orbit with a massive
dark companion, $M \gtrsim 2.3~M_{\odot}$ \citep{Cowley_1983}, LMC X-3
became the first example of an extragalactic stellar-mass black hole.

LMC X--3 is an unusual system straddling the boundary between
transient and wind-fed black hole binary systems (e.g.,
\citealt{Soria_2001}).  Its global characteristics, however, show that
its mass transfer is governed by Roche-lobe overflow, which places it
with the transients.  Although it exhibits extreme X-ray variability,
spanning roughly three orders of magnitude in luminosity, LMC X--3 is
unique among the transients because, with a few notable exceptions
\citep[e.g.][]{Smale_2012,Wilms_2001}, it remains in a disk-dominated
thermal state (approximately 90\% of the time).

In fact, the dominance of the soft thermal component and the relative
simplicity of LMC X--3's spectrum are the reasons \citet{Kubota_2010}
and \citet{Straub_2011} both chose the source to benchmark the
performance of accretion-disk spectral models.  Likewise, our group
has used the extensive thermal-state data available for LMC X--3 to
firmly establish the existence of a constant inner-disk radius in
black hole binary systems \citep{Steiner_2010}.

Recently, \citet{Smale_2012} have identified two long-duration ``low''
phases, each lasting several months, during which the system was in a
hard state.  This unusual behavior is likely related to the well-known
variability of the source on long timescales.  Initially, Cowley et
al. suggested a superorbital X-ray period of $\sim198$ (or possibly
$\sim99$) days, but subsequent studies of both optical and X-ray
variability find a range of superorbital periodicities extending from
$\sim100-500$ days \citep{Brocksopp_2001, Wen_2006, Kotze_2012}.
There is no stable long period in this source, and the superorbital
variability appears not to be attributable to the precession of a
warped disk or related at all to the orbital dynamics of the system.
Instead, the long-term variability is likely produced by changes in
the mass accretion rate (\citealt{Brocksopp_2001}; hereafter B01).

Several groups have used the extensive data available for LMC X--3 to
search for connections between X-ray variability and optical
variability.  Notably, B01 find that the X-ray and optical light
curves are correlated, with an X-ray lag of 5--10 days, whereas
\citet{Cowley_1991} find a somewhat longer $\sim 20$-day lag.

A complementary body of work has focused on examining very fast
($\lesssim 10$s) variability in the hard state for other black-hole
binary systems (e.g., \citealt{Gandhi_2010, Durant_2011}).  In these
systems, a surprising subsecond anticorrelation between X-ray and
optical is commonly observed that is unlikely to be related to X-ray
reprocessing.  Rather, it appears that this behavior is related to the
presence of a jet.  In the case of LMC X--3, no jet is expected,
particularly for the thermal state relevant here.

In the standard $\alpha$-disk theory of accretion \citep{SS73}, one
naturally expects the signals between any two wave bands to be
correlated with a time delay that corresponds to the viscous time for
a parcel of gas to travel from the outer emission region (longer
wavelength band) in the disk to the inner (shorter wavelength band).
Let us consider an annular region with an outer radius $R$ and a gap
in radius $\Delta R$ between the two zones that are of interest in
this paper, namely, the regions of optical and X-ray emission.  Since
the smaller radius in this case is negligible compared to $R$, the
viscous time for gas with viscosity $\nu$ to traverse the distance
$\Delta R \sim R$ is simply,
\begin{equation} 
t_{\rm visc} = R^2/\nu.
\label{eq:tvisc} \end{equation}

In this paper, we link the optical variability of LMC X--3 to its
X-ray variability by using a wealth of optical and X-ray data.  More
specifically, we employ two extensive optical / infrared (OIR) data
sets and, in the X-ray band, we combine data collected daily by
\rxte's All-Sky Monitor (ASM) with high-sensitivity data obtained in
many hundreds of pointed observations using \rxte's Proportional
Counter Array (PCA).

\section{Data}

Our principal OIR data set was obtained using the ANDICAM
\citep{Bloom_2004} on the \smarts\ 1.3~m telescope
\citep{Subasavage_2010}.  ANDICAM is a dual-channel imager, which we
used to obtain pairs of images of LMC X--3 in $B$, $V$, $I$, or $J$.
The data were reduced and calibrated with respect to several field
stars, as described in \citet{Jerry_LMCX3}.

The ANDICAM data were supplemented by $B$ and $V$ data kindly supplied
by C.\ Brocksopp and described in B01.  These data are a subset from a
larger sample obtained using the 0.91~m Dutch Telescope of the
European Southern Observatory between 1993 and 1999 during a total of
16 different observing runs.  Our full data set is derived from 470,
514, 476, and 440 \smarts\ observations in $B, V, I,$ and $J$,
respectively, and 356 and 685 additional ESO observations in $B$ and
$V$ (B01).  We employ just $\sim 75\%$ of the B01 data set, because we
exclusively consider those data collected after 1996 March 7 (MJD $\geq 50149$)
when \rxte\ operations entered a mature phase.

Our \rxte X-ray data set is derived from both ASM and PCA
observations.  The calibrated ASM data using the full bandwidth
(2--12~keV) and 1-day time bins were retrieved from the website
supported by the ASM/{\it RXTE}
team\footnote{http://xte.mit.edu/ASM\_lc.html}.  The PCA data, derived
from 1598 observations made using the PCU-2 detector, were all reduced
and analyzed following closely our standard methods (e.g.,
\citealt{Steiner_2010}), but in this case the data were partitioned in
somewhat finer time bins, using continuous exposure intervals with a
mean of 2~ks (and range 300s to 5000s).  Our goal here is relatively
modest, namely, to derive reliable estimates of flux and construct an
X-ray light curve.

\begin{figure}
{\includegraphics[clip=true, angle=90,width=9.cm]{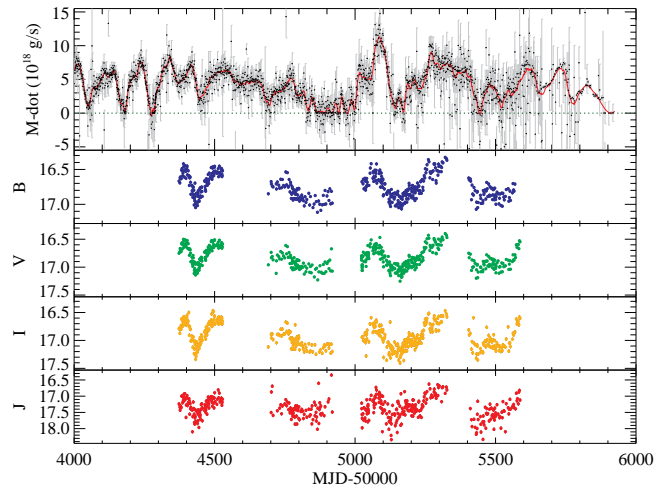}}
\caption{\smarts\ OIR and \rxte\ X-ray (ASM + PCA) light curves.  The
  light curve amplitude (top) is expressed in units of $\mdot$, as
  discussed in \S~\ref{subsection:mdot}. The red curve shows our
  adopted, smoothed light curve.} \label{fig:lc}
\end{figure}

The \smarts\ OIR light curves and the contemporaneous ASM + PCA X-ray
light curve are shown in Figure~\ref{fig:lc}.

\subsection{Unifying PCA and ASM Data Sets}\label{subsection:mdot}

In order to most fully capture the X-ray variability of LMC X--3, we
combine the ASM and PCA data sets into a single, unified light curve.
Given the complementary cadence and sensitivity of the two
instruments, this approach results in a light curve that for our
purposes is distinctly superior to using only one of the light curves
singly.  The method we adopt here for combining the light curves does
not critically affect our final results; e.g., the character of our
results are unaffected if we use the ASM data alone.

We choose to standardize the two data sets by determining for each ASM
and PCA observation the mass accretion rate onto the black hole,
$\mdot$.  We make this choice because LMC X--3 is nearly always in the
thermal state, a state in which the stable inner-radius results in a
constant efficiency (depending only on the spin parameter), and
$\mdot$ is therefore simply proportional to the bolometric luminosity
emitted by the disk.  The mass accretion rate determined from PCA
spectral fits can the be compared to the 2--12~keV PCA flux, providing
a scaling relationship that is then used to determine $\mdot$ values
for the 2--12~keV flux measurements from ASM data, which cannot be fit
for bolometric disk luminosity.  Fortunately, when LMC X--3 does
occasionally enter the hard state its luminosity is so low that $\mdot
\approx 0$ is a quite reasonable approximation for the purposes of
this study.

\begin{figure}
{\includegraphics[clip=true, angle=90,width=9.cm]{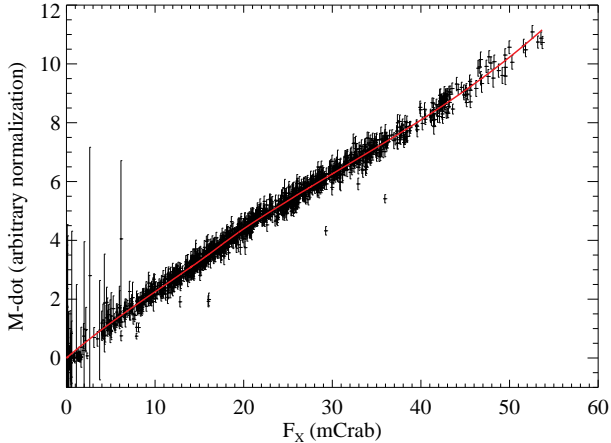}}
\caption{Spectral fitting results for the $\sim$1600 PCA spectra showing $\mdot$ as
  a function of 2--12 keV X-ray flux.  The \lowess\ fit we adopt is
  overlaid in red.  Results for 6\% of the full PCA sample are omitted
  from the figure either because the spectral fit used to derive
  $\mdot$ was statistically poor or it provided no constraint on
  $\mdot$.}\label{fig:mdot}
\end{figure}

We first determine $\mdot$ for the PCA data by fitting the $\sim 1600$
PCU-2 spectra to a standard disk model that accommodates the presence
of a weak Compton component.  The complete spectral model in {\it
  XSPEC} notation \citep{XSPEC} takes the form: {\sc
  tbabs}$\times(${\sc simpl}$\otimes${\sc kerrbb2}).  The disk
component \kerrbbtwo\ \citep{McClintock_2006,KERRBB} fits for $\mdot$
using fixed -- and here, approximate -- values for black-hole mass,
inclination and distance.  \simpl\ models the Compton power law
\citep{Steiner_simpl} and \tbabs\ the low-energy absorption
\citep{TBABS}.  Each PCA spectrum thus delivers a value of the
accretion rate\footnote{The other fit parameter of the disk component,
  spin, is allowed to vary, but is distributed with modest scatter
  about a fixed value.  A separate paper is forthcoming on the spin of
  LMC X--3 \citep{Steiner_LMCX3spin}.}  At the same time, we compute
the {\em observed} PCA flux $F_X$ from 2--12 keV relative to the
standard flux of the Crab in the same band \citep{Toor_Seward}.  As
illustrated in Figure~\ref{fig:mdot}, the relationship between $\mdot$
and $F_X$ obtained in this way is well defined and tightly
constrained\footnote{Roughly 0.7\% of the data, 11 data points, fall
 $>5\sigma$ below the curve.  These exceptional spectra show an
  unusually strong Compton component, more than an order of magnitude
  brighter than is typical for the disk-dominated states of LMC
  X--3.}.

We likewise converted the 2--12 keV ASM count rates into fluxes
relative to the Crab (1 Crab = 75 ASM c~s$^{-1}$) and then
interpolated the relationship derived for the PCA in order to
establish the relation between flux and $\mdot$ for the ASM.  (To
accommodate negative ASM count rates, we used an extrapolation of the
low-flux PCA data rather than adopting a floor value for $\mdot$.)

Having established a common scale for the two detectors, we then used
the \lowess\ nonparametric smoothing algorithm \citep{LOWESS} to
derive a representation of the true, underlying light curve (which is
at all points locally determined).  Because the errors are grossly
different for the PCA and ASM data, we achieve appropriate weighting
via Monte-Carlo randomization: A \lowess\ curve fit is derived for
each of 1000 random realizations and the median curve is taken as the
most representative fit.  Here, we used a third-order \lowess\ curve
fit and adopted a window length of approximately one month, a choice
motivated by the X-ray autocorrelation time (\S~\ref{section:xcors}).
The precise settings used in fitting the data only affect our results
cosmetically.  For example, with a window length of 3 days, we
obtained the same - though noisier - results.

\section{Correlated Variability}\label{section:xcors}

With the unified X-ray light curve now in hand (i.e., $\mdot (t)$), we
use the unsmoothed data and examine the relationship between X-ray and
OIR variability.  Specifically, we assess the correlated variability
between $\mdot$ and the OIR bands as a function of time lag using the
discrete correlation function (DCF) of \citet{EK88}, with all
observation times referred to the Solar System barycenter.  A DCF is
derived from a pair of time series, $A_i$ and $B_j$, by first
establishing a set of lags $\tau$ and cross-fluxes $\Upsilon$, which
are calculated across all paired time differences amongst the data:
\begin{equation}
\Delta t_{ij} = t_i - t_j, 
\label{eq:tlag}
\end{equation}
\begin{equation}
\Upsilon_{ij} = \frac{(A_i - \bar{A})(B_j-\bar{B})}{\sigma_A\sigma_B},
\label{eq:xflux}
\end{equation}
where $\bar{A}$ and $\bar{B}$ are the means of series $A$ and $B$,
respectively.  Similarly, $\sigma_A$ and $\sigma_B$ are the standard
deviations of $A$ and $B$.  The DCF for time lag $\tau$ is then
\begin{equation}
{\rm DCF}(\tau) = \frac{1}{N(\tau)}\sum_{\Delta t_{ij} \in \tau}\Upsilon_{ij},
\label{eq:dcf}
\end{equation}
where $N(\tau)$ is the number of $\Delta t_{ij}$ elements in the bin
$\tau$.

\begin{figure}
{\includegraphics[clip=true, angle=90,width=9.cm]{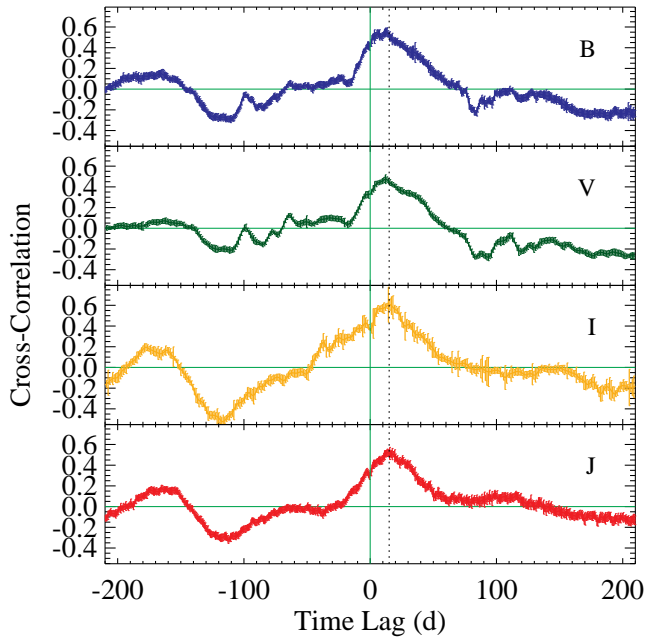}}
\caption{Cross-correlations between X-ray and four OIR bands.  The
  vertical dotted line at 15 days is meant to guide the eye in
  assessing the $\approx$~2 week X-ray lag.  Random realizations have
  been used to assess the error for each cross-correlation, which is
  the source of scruff on the curves. }\label{fig:xcors}
\end{figure}

We convert the OIR data into fluxes and then standardize all data sets
to unity variance and zero mean and then compute a DCF between each
OIR light curve and the (unsmoothed) $\mdot$ time series
(\S~\ref{subsection:mdot}), binning the time lag data into $\sim$1~day
intervals.  Given this sampling and given data that span a decade,
correlated variability is discernible over timescales ranging from
days to years.  As shown in Figure~\ref{fig:xcors}, the X-ray time lag
for the strongest features in the cross correlation of $\mdot$ with
the OIR flux in each individual band is $\approx 2$~weeks.  This value
is intermediate between those found by B01 and \citet{Cowley_1991}.  A
possible explanation for the differences is the lag's luminosity
dependence discussed in \S~\ref{section:discussion}.

The lag of the X-ray flux is naturally explained as a consequence of
the viscous time delay for a density perturbation to propagate from
the outer disk, the location of peak OIR emission, to the vicinity of
the black hole where the X-rays are produced.

\begin{figure}
{\includegraphics[clip=true, angle=90,width=9.cm]{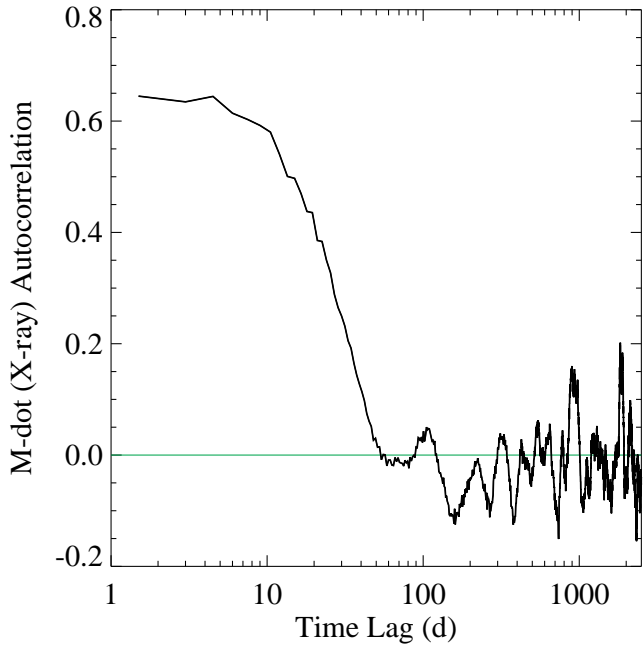}}
\caption{X-ray autocorrelation.  The typical $\approx 20$-day
  ``pulse-width'' response of the X-ray light-curve shows that
  significant changes in the accretion flow require several weeks to
  materialize.  This timescale is responsible for the breadth of the
  lag features in Fig.~\ref{fig:xcors}. Errors are shown but are
  generally comparable to or smaller than the
  linewidth.}\label{fig:xauto}
\end{figure}

Why are the cross-correlation peaks so broad?  An examination of the
X-ray autocorrelation (Figure~\ref{fig:xauto}) suggests an answer.
The autocorrelation timescale over which the X-ray source brightens or
dims, several weeks, is similar to the typical lag time of $\approx
2$~weeks.  This is not surprising given that the timescale for the
inner disk to change from faint to bright -- the autocorrelation time
-- should be commensurate with the time required for a parcel of gas
to travel from the outer disk to the center, i.e., the viscous time.

On longer timescales, first at $\sim100$ days and $160$ days,
other features appear in the autocorrelation function
(Figure~\ref{fig:xauto}).  As evidenced by alternating
positive/negative harmonic peaks, the anti-correlation feature at 160
days, e.g., is likely to be at least partially related to radiative or
mechanical feedback in the system, whereby periods of strong accretion
inhibit the accretion flow onto the outer disk.  This is also evident
in the strong anti-correlation signature between OIR and X-ray bands
at $\sim -120$d lag (where the optical behavior now follows after the
X-ray).

In the next section, we develop a model for the accretion-driven,
$\approx 2$~week time lag in the context of a simplistic physical
picture of the disk-star system in LMC X--3.

\section{The Accretion Model}\label{section:accmodel}

\begin{figure}
{\plotone{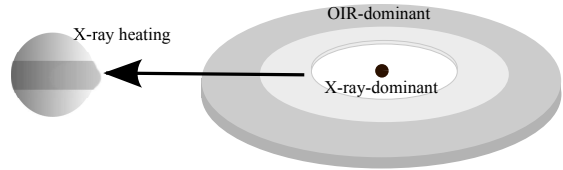}}
\caption{A schematic representation of our model for the binary
  system.  The companion star on the left is shown heated by the
  central X-ray source at high latitudes and shielded by the disk in
  the equatorial region.  The outer and inner regions of the disk
  ({\em not} to scale) produce respectively OIR and X-ray emission.
}\label{fig:schem1}
\end{figure}

Our purpose is to model the complex OIR light curves of LMC X--3 as a
blend of three components of emission attributable to (1) the
companion star; (2) a multitemperature thermal disk; and (3) the
regions in the disk and star where X-rays are reprocessed.  (We need
not consider synchrotron emission here, as there is no evidence for a
jet in LMC X--3, unlike many other sources.) Our model for the system
is shown schematically in Figure~\ref{fig:schem1}.  We approximate the
net emission $F(t)$ in a given OIR band $i$ as a sum of the three
components:
\begin{equation} F_{i}(t) = X_i(t) + S_i(t) +  H_i(t), 
\label{eq:flux} 
\end{equation}
where $X_i(t)$ is the OIR disk emission which relates to the later
X-ray emission (or $\mdot$) at time $t+t_{\rm visc}$; $S_i$ is the
direct flux from the star plus any steady component of disk light; and
$H_i$ is the reprocessed emission from the outer disk and companion
star.

\subsection{Stellar Component}\label{subsec:stellarmod}

The simplest term, $S_i$, is a constant plus sinusoidal terms that
approximate the asymmetric ellipsoidal variability of the tidally
distorted star: 
\begin{equation} S_i(t) \equiv C_i + c_{i} \left[ {\rm sin}(
    4\pi \phi - \pi/2) + \epsilon e(\phi) \right],
\end{equation}\label{eq:ellipse} where phase $\phi \equiv (t-t_0)/P$
and \begin{equation}
 e(\phi) \equiv {\rm max}\left(0, {\rm sin}(4\pi \phi + \pi/2)\right) \times \left\{ 
  \begin{array}{l l} 
    1,   & \quad {\rm sin}(2\pi \phi + \pi/4) > 0,\\
   -1,   & \quad {\rm sin}(2\pi \phi + \pi/4) \leq 0
\end{array}  \right.
\label{eq:asym}
\end{equation}
allow for the difference in the amplitudes of the two minima.  Both
$\epsilon$ and $c_i$, the normalizations respectively of the
asymmetric and sinusoidal terms, are free parameters, as are $t_0$,
$C_i$, and $P$.

\subsection{Disk Component}\label{subsec:diskmod}

The connection of the term $X_i(t)$, the variable OIR component of
disk emission, to $\mdot$ is obvious.  However, $\mdot$ was derived
for the inner X-ray-emitting portion of the disk, which is located far
from the optical-emitting region, and the mass transfer rates may be
different in the two regions.  Therefore, in order to compute the OIR
disk emission from $\mdot$, our model must include a description of
how matter flows from the outer disk to the inner region.  This is
established via a transfer function which maps between X-ray and OIR
regimes.  We envision the disk described with such a model as a series
of concentric rings.  The accretion flow for each individual ring is
approximately steady-state, but the steady-state solution varies from
ring to ring.  The order of the rings are fixed, but they are allowed
to stretch or compress in width during inflow.

Meanwhile, the observed, viscosity-induced time delay of the X-ray
flux relative to the optical will have some dependence on the mass
supply of the individual rings.  In our generic prescription, we
incorporate this dependence as a free power-law scaling on the time
delay, which depends on $\mdot$, a choice that is explained further in
Section~\ref{section:discussion}.  We likewise assume that the optical
emission has a power-law dependence on $\mdot$:
\begin{equation}
X_i(t) \equiv a_{i} \times \left[\mdot(t+\delta t(t)) /10^{18}g\;s^{-1}\right]^\beta,
\label{eq:f_lag}
\end{equation}
with $a_{i}$ and $\beta$ as free parameters and $\delta t(t)$ is the
viscous lag at time $t$, defined as
\begin{equation}
\delta t (t) \equiv \Delta_i \left(\frac{\mdot(t+\delta t(t))}{< \mdot >} \right)^\psi,
\label{eq:lag}
\end{equation}
where $\Delta_i$ and $\psi$ are free parameters.  A slightly modified
prescription for the accretion emission is considered in
Appendix~\ref{section:model2}.
Finally, we allow for a potential scaling error in our conversion from
X-ray flux to $\mdot$
by introducing a floating offset $a_X$ which modifies the
zero-point of $\mdot$. For simplicity, rather than introducing a new
variable, $\mdot$ will be understood to contain an implicit free
zero-point, which is determined in the fit.

\subsection{Reprocessed Emission}\label{subsec:xheatmod}

The X-ray heating by the central source, which occurs on a timescale
of seconds, is treated as instantaneous.  That is, the heating term
$H_i(t)$ responds directly to changes in $\mdot(t)$ (zero time lag).
We arbitrarily parameterize the dependence of $H_i(t)$ on $\mdot$ as a
broken power law.  This choice allows for variable shielding by the
disk as the X-ray luminosity varies.

The heating term contains two elements: The primary term, which has no
orbital phase dependence, describes X-ray heating of the outer disk;
and the secondary term, $Z(t)$, describes X-ray heating of the
companion star.  $Z(t)$ depends on phase and is maximum at superior
conjunction of the star.  We assume that $Z(t)$ varies sinusoidally
and that its normalization constant $h_i = h$ is the same for all of
the OIR bands.  This approximation is valid because for the B-type
companion star \citep{Cowley_1983} the OIR bands all lie in the
Rayleigh-Jeans part of the spectrum.

\begin{equation}
  H_i(t) \equiv  h~c_{i}\left(1+Z(t)\right) \left\{ 
      \begin{array}{l l l l}
        (\mdot(t))^{\gamma_1},  &  \mdot(t) < \dot{m}_{\rm break}, \\
        \dot{m}_{\rm break}^{\gamma_1} \left(\frac{\mdot(t)}{\dot{m}_{\rm break}}\right)^{\gamma_2},  & \mdot(t) > \dot{m}_{\rm break},
      \end{array}  \right.
    \label{eq:xheat}
  \end{equation}
where 
\begin{equation}
Z(t) = \eta/2\left(1+{\rm sin}(2\pi\phi-\pi/2)\right),
\label{eq:shadow}
\end{equation}
so that $Z(t)$ varies between 0 and $\eta$ each orbit, where $\eta$,
$h$, $\gamma_1$, $\gamma_2$, and $\dot{m}_{\rm break}$ are fit
parameters.

Our full model consists of a total of 27 free parameters (plus two
additional calibration parameters to align the B01 and \smarts\
datasets).  There are 16 ``chromatic'' parameters that contain a
subscript $i$ (4 each per OIR band) and there are 11 ``gray''
parameters that are independent of wavelength (e.g., the orbital
period $P$).  Correlations amongst the fit parameters are discussed
and shown in Appendix~\ref{append:correlations}.

\section{Results}\label{section:results}

The model has been implemented in {\em python} and applied at once to the
composite OIR data set.  All fits are computed using the Markov Chain
Monte-Carlo (MCMC) routine {\sc emcee-hammer} \citep{emcee} in a Bayesian
formalism.  Because our model is an oversimplification of complex real
processes, we expect the quality of
the fits to be limited by systematic effects.  We therefore adopt 10\%
and 20\% (systematic) error bars for the optical and IR data,
respectively, and we ignore the much smaller measurement errors.  These
round values were chosen by making a preliminary fit to the data and
assessing the rms fit residuals.  Because the uncertainties were
estimated in this utilitarian way, any goodness-of-fit statistics -such
as $\rchinu$ - should be interpreted with reserve.

We use a flat prior (i.e., uniform weighting) on the period; for all
other parameters, we use informed priors whenever one is evident.
Otherwise, we default to flat priors on the logarithm of normalization
terms (i.e., scale-independent weighting), and flat priors on shape
parameters.  The fitting results, which are based on our analysis of
all the OIR data, are presented in Table~\ref{tab:fits}.  MCMC
directly computes the posterior probability distribution of a model's
parameters, but it does not specifically optimize the fit quality.
Therefore, the best-fitting value of $\rchinu$ given in
Table~\ref{tab:fits} has been obtained via other standard optimization
methods (e.g., Levenberg–Marquardt, downhill-simplex, etc.), while
taking the MCMC results as a starting point.

To achieve our fits via emcee-hammer, we used 500 walkers with 15000
steps apiece for a total of 7.5 million MCMC samples.  Our analysis
and results are based on the final 2.5 million samples.  Convergence
has been diagnosed following \citet{Gelman_Rubin} using a stringent
criterion of $\tilde{R} < 1.1$.  For our analysis, across all
parameters, $1.049 < \tilde{R} < 1.085$.

\begin{figure*}
{\includegraphics[clip=true, angle=90,width=15.85cm]{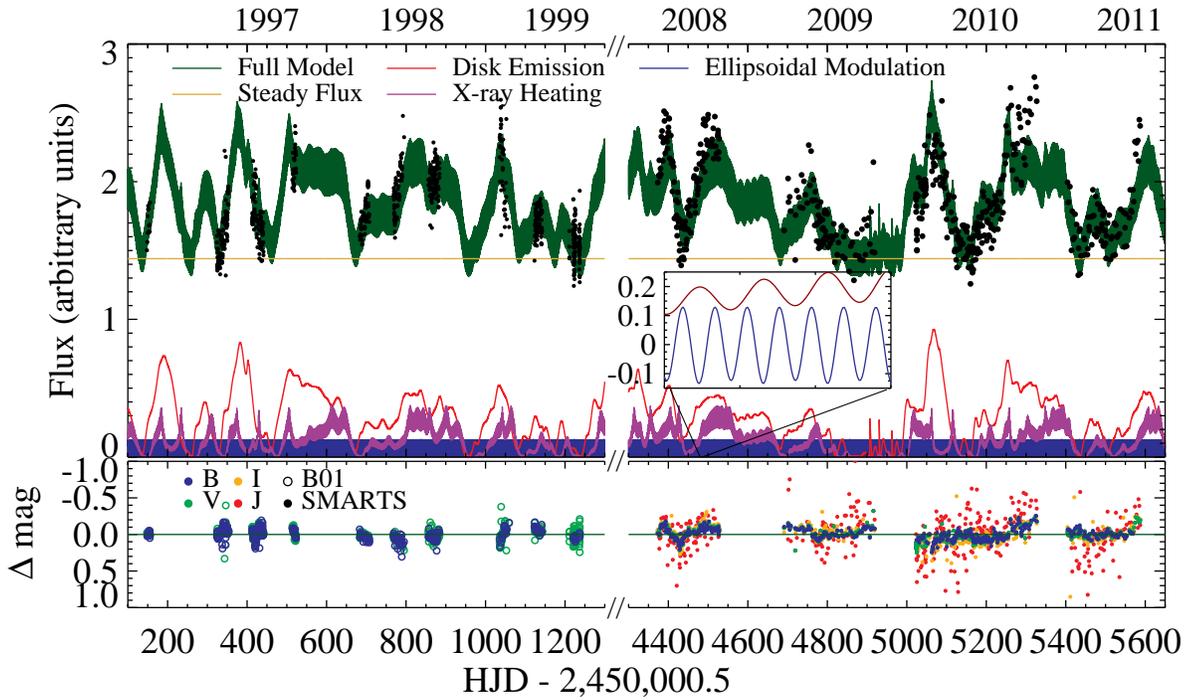}}
\caption{The model of the observed light curve and its three
  components for the band with the most data, $V$.  The composite
  model is shown in dark green; the accretion emission derived using
  the X-ray data in red; and the ellipsoidal and X-ray-heating
  components are shown in blue and purple, respectively.  The inset is
  a blowup showing the ellipsoidal and X-ray heating terms for several
  orbital cycles.  The fit residuals are shown in the lower panels.
}\label{fig:panels}
\end{figure*}

In Figure~\ref{fig:panels}, we show for the $V$-band light curve (the
band with the most data) the best-fit model broken down into its
component contributions.  The figure makes clear that the total flux
is dominated by the constant component of stellar light, while the
disk and X-ray-heating component contributions are at most $\sim 1/2$
and $\sim 1/4$ as large, respectively (as determined in the fit).  The
{\em maximum} contribution of X-ray heating to the total stellar flux
is $h{{\dot{m}}_{\rm break}}^{\gamma_1} \sim 0.16$\%.  Meanwhile, the
contribution to the stellar flux due to ellipsoidal variability is
$c_i/C_i \approx 5-10\%$ (Section~\ref{subsec:stellarmod}), while the
asymmetric term is a factor $\sim 10$ smaller still.  B01 conclude
that the optical variability of LMC X-3 is due to viscous disk
emission rather than X-ray reprocessing, i.e., that reprocessing is of
secondary importance.  Our results support B01's conclusion.

Our fits indicate that as the X-ray flux varies over its full range that
the fraction of stellar light to the total light in both the $B$ and $V$
bands is $\sim 80\%\pm10\%$.  A consistent value of $\approx 70\%-90\%$
has been derived independently for a wide range of X-ray luminosities
using spectroscopic data (\citealt{Jerry_LMCX3}).

The value of the scaling index $\beta = 1.31\pm0.08$ (90\%
confidence), which relates the OIR flux to the time-lagged disk
emission $\mdot$ (Section~\ref{subsec:diskmod}), may indicate that the
rates of inflow at the outer and inner radii are not matched.  If so,
the implication is that mass is lost from the body of the disk,
especially at high values of $\mdot$.  This scenario is readily
explained by the action of disk winds (e.g.,
\citealt{Miller_2006_H1743, Luketic_2010, Ponti_2012, Neilsen_2012}).
Alternatively, as discussed in \S~\ref{section:discussion}, a standing
shock from the accretion stream could be responsible for the mismatch
in rates.

From the flux densities computed in Table~\ref{tab:fits}, very rough
constraints can be placed on the temperature of the star and of the
OIR-emitting region of the disk.  However, such calculations are
fundamentally limited by the accuracy of the absolute OIR flux
calibration, which is marred by 20--30\% zero-point calibration
differences between \smarts\ and B01 data sets.  We rely on the
\smarts\ calibration as our standard rather than that of B01 because
the former calibration is based on a larger sample of standard stars
that were observed more frequently.



  \begin{deluxetable*}{ccccccccccccccc} 
  \tabletypesize{\scriptsize} 
  \tablecolumns{      7}
  \tablewidth{0pc}  
  \tablecaption{Best-Fitting Model}
  \tablehead{\colhead{Param.} & \colhead{Global} & \colhead{B} & \colhead{V} & \colhead{I} & \colhead{J} & \colhead{Prior\tablenotemark{a}}}
  \startdata  

$P$ (d)                                                     &    $           1.704805\pm3\times10^{-6} $  & \nodata & \nodata & \nodata & \nodata & Flat \\
$\beta$                                                   &   $                                1.31 \pm 0.08 $  &  \nodata & \nodata & \nodata & \nodata &$N(1,0.5)$ \\
$\dot{m}_{\rm break}/{\rm max}(\mdot)$   &  $                    0.545^{+ 0.017}_{- 0.006} $  &  \nodata & \nodata & \nodata & \nodata & Log \\

$\psi$                                                      &  $                       0.24^{+ 0.03}_{- 0.04} $  &   \nodata & \nodata & \nodata & \nodata & Flat \\
$a_X (10^{18}~g\;s^{-1})$                           &  $                       0.08^{+ 0.11}_{- 0.03} $  &  \nodata & \nodata & \nodata & \nodata & Flat \\
$t_0-T_{0,{\rm Cowley}}$\tablenotemark{b} (d)                          &  $                    0.004^{+ 0.012}_{- 0.017} $  &  \nodata & \nodata & \nodata & \nodata & $N(0,0.01)$ \\
 $\Delta_{i}$ (d)                                        &  \nodata & $                                 15.5 \pm 1.1 $  & $                                 15.4 \pm 0.8 $  & $                                 16.4 \pm 1.1 $  & $                            16.^{+ 1.}_{- 3.} $  & Flat \\
$a_{i}  (\uJy) $                                          &   \nodata & $                                   17 \pm 3 $  &$                                   15 \pm 3 $  & $                          9.4^{+ 2.1}_{- 1.6} $  & $                          5.0^{+ 1.3}_{- 0.8} $  & Log \\
$c_{i}  (\uJy)$                                           &   \nodata & $                                   62 \pm 6 $  & $                                   46 \pm 4 $  & $                                   22 \pm 4 $  & $                             9^{+ 5}_{- 3} $  & Log \\
$C_{i}  (\uJy)$                                          &    \nodata & $                                 670 \pm 11 $  & $                                  547 \pm 8 $  & $                                  285 \pm 6 $  & $                                  146 \pm 4 $  & Log \\
$\Delta m_{\rm B}$\tablenotemark{c}       &    $                              0.293 \pm 0.012 $  & \nodata & \nodata & \nodata & \nodata & Flat \\
$\Delta m_{\rm V}$\tablenotemark{c}       &    $                    0.216^{+ 0.008}_{- 0.012} $  &\nodata & \nodata & \nodata & \nodata & Flat \\
${\rm log}_{10}(h)$                                   &    $                         -2.3^{+ 0.2}_{- 0.3} $  &\nodata & \nodata & \nodata & \nodata & Log on $h$\\
$\gamma_1$                                           &    $                                  1.9 \pm 0.3 $  &\nodata & \nodata & \nodata & \nodata & $N(1,0.5)$ \\
$\gamma_2$                                           &    $                         -3.4^{+ 0.7}_{- 1.0} $  &\nodata & \nodata & \nodata & \nodata & Flat \\
 $\eta$                                                    &     $                          0.7^{+ 0.3}_{- 0.2} $  &\nodata & \nodata & \nodata & \nodata & Flat \\
 ${\rm log}_{10}(\epsilon)$                       &     $                         -1.1^{+ 0.4}_{- 0.6} $  &\nodata & \nodata & \nodata & \nodata & Flat on $\epsilon$ \\

\hline
$\rchinu$\tablenotemark{d}     &   2426 / 2894 \\   


\enddata
\tablecomments{Best fit model and associated 90\% credible intervals. 
Of the 29 parameters, 16 are normalization terms to the component fluxes.}
\tablenotetext{a}{The shape of the prior assumed.  ``Flat'' indicates
  uniform weighting, and ``log'' refers to uniform weighting on the
  logarithm of the parameter. $N$($\mu$,$\sigma$) describes a Gaussian
  centered upon $\mu$ with standard deviation $\sigma$.}
\tablenotetext{b}{$T_{0,{\rm Cowley}}$ =  2445278.66 HJD \citep{Cowley_1983}. }
\tablenotetext{c}{Zero-point differences between the datasets of B01
  and \smarts.}  
\tablenotetext{d}{The goodness-of-fit is calculated
  using 10\% systematic uncertainty on the optical model and 20\%
  systematic uncertainties on the IR; the small measurement errors on
  the data contribute negligibly and were ignored.}
\label{tab:fits}
\end{deluxetable*}

\begin{figure}
{\includegraphics[clip=true, angle=90,width=9.cm]{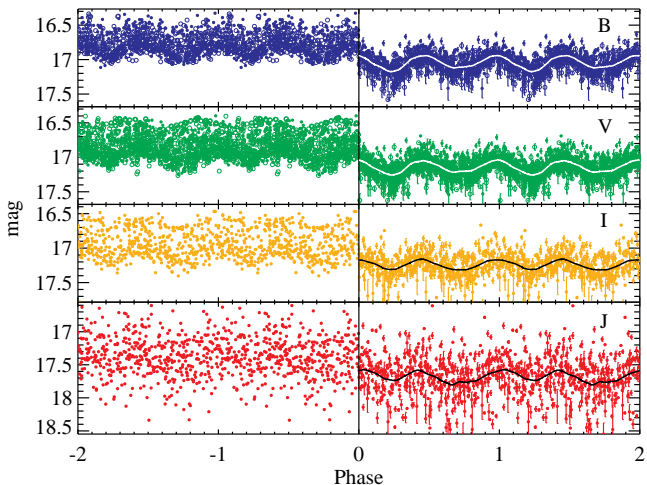}}
\caption{ {\it (left)} OIR light curves folded on the orbital period.
  A zero-point offset has been applied to bring the B01 data (open
  symbols) into alignment with the \smarts\ data (filled symbols).
  The shift is roughly 0.3 and 0.2 mag in $B$ and $V$ respectively
  (see Table~\ref{tab:fits}).  {\it (right)} OIR light curves obtained
 after removing the heating and disk-emission terms $H(t)$ and
  $X(t)$.  Error bars reflect an estimated 20\% uncertainty on the
  accretion ``noise'' which has been subtracted.  A {\sc lowess}
  smoothing of the light curve is overlaid on the data in each panel
  to guide the eye.  }\label{fig:denoised}
\end{figure}

In Figure~\ref{fig:denoised}, we show two versions of the OIR light
curves folded on LMC X--3's orbital period: On the left are the light
curves in their original form, and on the right are the filtered light
curves produced by removing the two nonstellar contributions, i.e.,
the $\mdot$-induced component $X(t)$ and the reprocessed X-ray
emission $H(t)$.  Plainly, the ellipsoidal variability is much more
apparent in the filtered light curves.

\begin{figure}
{\includegraphics[clip=true, angle=90,width=8.8cm]{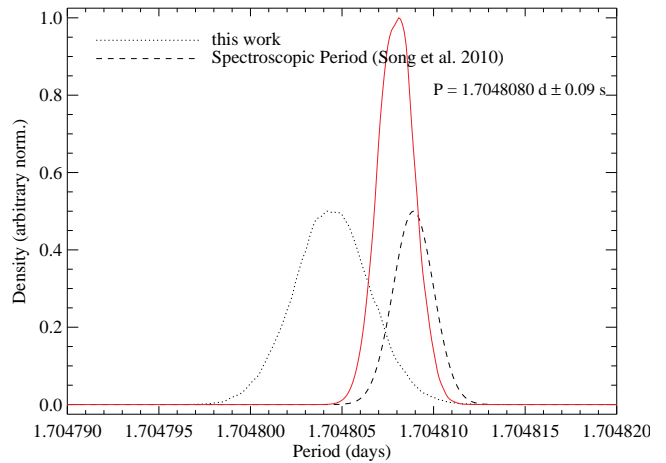}}
\caption{Our result for the orbital period alongside the definitive
  spectroscopic determination \citep{Song_2010}. The combined
  result is shown in red.}\label{fig:period}
\end{figure}

Because of the long baseline and abundance of data, the signal evident
from the filtered light curves (using the complete model) allows one
to determine the $1.7$-day orbital period to the remarkable precision
of two-tenths of a second.  (Prior to this work, the best
determination of the period has been $P=1.7048089 \pm
1.1\times10^{-6}$~d (\citealt{Song_2010}); our result is in good accordance
with theirs.)  The orbital phase, however, conforms to its
prior and is otherwise unconstrained.  The quality of our period
determination is illustrated in Figure~\ref{fig:period} where it is
shown to compare favorably with an independent determination based on
three decades of spectroscopic velocity data.  As indicated in the
figure, the two data sets jointly determine the period to a precision
of 90 milliseconds ($P = 1.704808 \pm 1\times 10^{-6}$~d).

The raw light curves, by comparison, cannot even deliver a unique
orbital period because of strong aliasing.  A Lomb-Scargle
\citep{LombScargle} search shows that the strongest false period is
favored 9:1 over the true period.  Moreover, the uncertainty in the
period determination is a factor $\sim 3$ worse than achieved using
the light curve model.

\section{Discussion}\label{section:discussion}

\subsection{Reprocessing and Self-Shadowing}

The intensity of the X-ray heating component in the OIR bands is a
surprisingly strong function of X-ray luminosity, scaling as $H_i
\propto \mdot^{1.9}$, up to a critical value of $\mdot \approx 6.2
\times 10^{18}$~g/s (roughly $\sim 50\%$ of the Eddington limit).
Above this luminosity, the heating signal drops off rapidly as
$\mdot^{-3.4}$ such that at the maximum X-ray luminosity ($\approx
L_{\rm Edd}$) its intensity is a factor of 10 below its peak.  A
likely explanation for the initial superlinear rise with luminosity is
that the flaring of the disk at higher luminosity increases the solid
angle of the reprocessing region, and that this effect dominates over
the self-shadowing by the disk.  Above $\sim 50\% L_{\rm Edd}$, the
shadowing dominates and the heating signal dwindles rapidly.


For the shadowing to be substantial without producing more OIR
emission than is observed, the flare in the disk's scale height must
move inward as the X-ray luminosity increases.  That is, we envision a
flared funnel-like region in the inner disk that contracts in around
the black hole as luminosity increases and as the flaring becomes more
pronounced.  The OIR emission from the reprocessing region remains
modest, even at the highest luminosities, because the reprocessing
excess is relegated to shorter wavelengths and emitted from within a
diminished area of the disk.  A prediction of this picture is that the
pattern of reprocessing shifting to smaller area and shorter
wavelength at growing luminosity should be observable in optical / UV
for those low-inclination systems which show substantial luminosity
variability.

As viewed from the X-ray source, the star subtends an angle of $\sim
37\degr$ (ignoring disk obscuration).  The corresponding fraction of the
X-ray heating signal that is reprocessed in the face of the star is obtained from the fit:
$\eta/(1+\eta) \approx 40\%$ (Equations~\ref{eq:xheat},\ref{eq:shadow}).   
 The reprocessed emission from the star can be
distinguished from the reprocessed disk emission because the former is
modulated at the orbital period.  For
simplicity, the fraction of the total reprocessed emission contributed
by the star is treated as a constant, free parameter.


\subsection{Wavelength-Independence of the Lag}

In canonical thin accretion disk theory \citep{SS73}, the wavelength
of peak emission from the disk scales as $\lambda_{\rm max}(R) \propto
R^{3/4}$, whereas for irradiation-dominated disks the peak wavelength
scales roughly as $\lambda_{\rm max}(R) \propto R^{3/7}$
\citep{Chiang_Goldreich_1997}.  On this basis, one expects a factor
$\sim 2$ variation in the viscous time lag across the OIR bands.
However, as evident from Table~\ref{tab:fits} and from inspection of
Fig.~\ref{fig:xcors}, the OIR time lag is essentially independent of
wavelength from the $B$-band ($\approx4400$~\AA) to the $J$-band
($\approx12,500$~\AA)\footnote {The wavelength-independence of the lag
  has been verified by examining sub-intervals of the data as well.}.

The constancy of the viscous time lag with wavelength implies that the
OIR signal arises from a single radius in the disk. This result is
incompatible with any simple disk theory for which temperature falls
off monotonically with radius, unless one imagines locating the radius
of OIR emission at the disk's outer edge.  However, the outer edge is
obviously ruled by the data because the entire disk would then be so
hot that it would outshine the star.

\subsection{Shocks and Hotspots in the Outer Disk}

\begin{figure}
\plotone{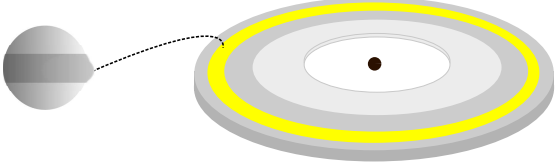}
\caption{The OIR lag signal is dominated by a bright ring (yellow) in
  the outer disk at which the accretion stream (dashed line) shocks into the disk.
  This ring may coincide with the circularization radius, where the
  angular momentum of the accretion stream (dashed line) matches the
  disk's Keplerian angular momentum.}\label{fig:schem2}
\end{figure}

Instead, the constancy of the time lag with wavelength is naturally
explained by the presence of a ``hot spot'' or similar structure
within the outer disk that is bright enough to dominate the OIR lag
signal.  Doppler modulation tomographic maps of accreting binary
systems show clear evidence for such features in other sources (e.g.,
\citealt{Steeghs_2004, Calvelo_2009, Kupfer_2013}).  An excellent
example is provided by the map of the black hole binary A0620--00,
which reveals bright spots embedded in a hot ``ring'' in the disk
(dominated by two large crescent structures; \citealt{Neilsen_2008}).
The ring is located at 45\% or 60\% of the L1 radius ($R_{\rm L1}$)
depending on whether one assumes, respectively, a Keplerian or
ballistic trajectory for the gas stream.  This range in radius exactly
brackets the {\em circularization radius} of the system, the radius at
which the angular momentum of the tidal stream matches the angular
momentum of the gas in the disk.  The circularization radius $R_{\rm
  C}$ is located at 50\%~$R_{L1}$ for A0620--00, while the truncation
radius of the outer disk is predicted to be located at $\sim
80-90$\%~$R_{L1}$ \citep{FKR}.  A schematic diagram of a system like
A0620--00 is shown in Figure~\ref{fig:schem2}.

The temperatures of these bright spots and rings are poorly constrained
in quiescent black-hole binaries, and they are essentially unconstrained
in active systems.  In order to crudely gauge the temperature of such a
spot, we look to the recent work by \citet{Kupfer_2013} where for a
cataclysmic variable (CV) system they have measured $T_{\rm hotspot}
\approx 30,000$~K.  Such hot spots are attributed to shocks produced by
the tidal stream of gas from the companion star impinging on the disk.
A strong shock (Mach number $\mathcal{M} > 10$) can result in up to a
hundredfold jump in temperature in the post-shock gas.

If the mass flow rate in the stream $\dot{m_{\rm s}}$ is sufficiently
large, a shock that traverses inward from $R_{\rm out}$ will
eventually stall at the circularization radius $R_C$.  For simplicity,
we assume that the flow is purely ballistic, i.e., that the velocity
is unaffected by $\dot{m_{\rm s}}$, and that the hotspot luminosity
increases as $\dot{m_{\rm s}}$ increases.  We now mention an
alternative explanation for the superlinear scaling between OIR and
X-ray emission ($\beta = 1.3$), which in Section~\ref{section:results}
we attributed to disk winds.  Here, where the OIR emission originates
in a shock, we instead envision that this nonlinear relationship could
plausibly result from changes in the structure or strength of the
shock caused by variations in $\dot{m_{\rm s}}$.

\subsection{Estimating $\alpha$-viscosity}

The cooling time for the shock-heated gas is negligibly short compared
to the viscous timescale.  Therefore, the stability and brightness of
these structures are related to the instantaneous value of $\dot{m_{\rm
s}}$.  The ring itself must be relatively narrow owing to the prompt
cooling.  The cooled gas relaxes to the disk profile, and then it
proceeds inward to the center on the viscous timescale.

The outermost disk is always dominated by gas pressure so that the scale
height at each radius is given, roughly, by the ratio of the sound speed
in the midplane, $c_s=\sqrt{kT/\mu m_p}$, to the Keplerian velocity in
the disk.  Using the $\alpha$ disk approximation for disk viscosity
\citep{SS73},
\begin{equation}
\nu = \alpha c_s z,
\label{eq:alphadisk}
\end{equation}
where $z$ is the scale height of the disk.  Then,
\begin{equation}
\nu \equiv \alpha c_s^2 \frac{R^{3/2}}{\sqrt{G M}},
\label{eq:visc}
\end{equation}
and the viscous timescale can be written as 
\begin{equation}
t_{\rm visc} \approx \sqrt{G M R} \left(\frac{\mu m_p}{\alpha k T}\right).
\label{eq:tviscx}
\end{equation}

In what follows, recall that we are approximating the evolving disk as a
superposition of independent, steady-state solutions
(Section~\ref{subsec:diskmod}).  We make two other assumptions, which
are eminently reasonable: (1) The disk is strongly illuminated; and (2)
$\dot{m_{\rm s}}$ alone determines the X-ray and OIR emission.  Building
on these assumptions and our measured value of the time lag, we now
derive a rough estimate of $\alpha$ in the outer disk.  We first
consider the conditions in the disk and the location of the
OIR-emitting hotspots.

The strong X-ray irradiation of the disk will modify its structure at
large distances from the central source, i.e., $R > 10^{10}$~cm
(\citealt{Hartmann_book}).  Meanwhile, the midplane temperature, being
less sensitive to irradiation, will tend to follow the viscous
dissipation profile, $T_{\rm m} \propto R^{-3/4}$.  However, the
temperature of the irradiated surface layer of the disk will fall off
much more slowly, as $T_{\rm s} \propto R^{-3/7}$
\citep{Chiang_Goldreich_1997}, or even as slowly as $T_{\rm s} \propto
R^{-1/3}$ \citep{Fukue_1992}.  To our knowledge, the vertical
temperature gradient in strongly irradiated disks has not been given
serious consideration.  Generally, $T_{\rm s}$ and $T_{\rm m}$ in the
outer disk, determined from irradiation and viscous dissipation,
respectively, are comparable and within a factor of a few of one
another, and it is reasonable to approximate the vertical temperature
profile as isothermal (e.g., \citealt{Cunningham_1976, King_1998}).
For the average mass accretion rate onto LMC X--3, we estimate for any
reasonable set of disk parameters that to within a factor $\sim3$ the
midplane temperature $T_{\rm C} \sim 10^5$~K at $R_{\rm C}$.

For this temperature and radius, using
Equations~\ref{eq:alphadisk}-\ref{eq:tviscx} we find to order of
magnitude that $\alpha \sim 0.5$.  Our result agrees well with other
measurements of $\alpha$, particularly those obtained for CVs
(\citealt{King_2007}, and references therein).  The most reliable of
these other measurements, like our estimate for LMC X-3, were derived
for irradiated outer disks.

\subsection{$\alpha$ Scales with Luminosity}

A surprising outcome of our model and analysis is the positive
luminosity scaling of the time lag ($\psi \approx 0.25$; Eq.~\ref{eq:lag}).
Recalling that $t_{\rm visc} \propto \dot{m}_X^{\psi}$,
Eq.~\ref{eq:tviscx} gives $\alpha t_{\rm visc} \propto T^{-1} \propto
\dot{m}_X^{-1/4}$.  For a constant value of $\alpha$, one expects to
find $\beta = -0.25$, whereas we find a positive value of $\beta$.  As a
bottom line, we find that the viscosity parameter varies with luminosity
(or equivalently, mass accretion rate) as $\alpha \propto
\mdot^{-1/2}$.   An inverse scaling such as we predict has in fact, already been
suggested for LMC X--3 based purely on changes in its disk spectra as
the luminosity varies \citep{Straub_2011}.

If this scaling is related to the increase in temperature which
results from a higher $\mdot$, then $\alpha$ may be lower in the inner
disk\footnote{On the other hand, this dependence cannot be {\em
    purely} due to an inherent scaling with temperature, i.e. $\alpha
  \propto T^{-2}$, without the $\alpha$-disk interpretation breaking
  down.  A weaker scaling with temperature would be compatible.}.
However, GRMHD simulations indicate that $\alpha$ increases as radius
decreases, at least for the innermost disk where MHD turbulence is
dominant \citep{Penna_2013}.  Of course, there is no reason to expect
viscosity to be constant over the disk.  In particular, the viscosity
may depend on other factors such as the disk's density, temperature or
magnetization.

\citet{Bai_Stone_2011} provide one possible explanation for the scaling
relation we find relating $\alpha$ and luminosity.  These authors find
an inverse scaling between $\alpha$ and the ratio of gas-to-magnetic
pressures, $P_{\rm gas}/P_{\rm mag}$ (a net quantity usually referred to
as ``plasma $\beta$'').  Given the scaling, $\alpha \propto P_{\rm
mag}/P_{\rm gas}$, if $P_{\rm mag}$ is sufficiently insensitive to the
accretion flow in the outer disk, then one would expect that as the mass
transfer rate ($\dot{m}$) varies, when $\dot{m}$ increases, $P_{\rm
gas}$ will too.  A more specific prediction would be beyond scope of
this work, but we note qualitatively that this effect very naturally
gives rise to an inverse relationship between $\alpha$ and mass
accretion rate, as required by our fit.

\section{Conclusions}\label{section:concs}

Motivated by a roughly two-week time-lagged correlation between the
OIR and X-ray light curves of LMC X--3, we develop a new method of
analysis that is applicable to active X-ray binary systems.  We model
the OIR emission deterministically as a combination of accretion
emission, X-ray reprocessing, and stellar emission.  The components
driven by accretion are computed using the X-ray light curve.  The
model allows accretion signals to be filtered out of the OIR light
curves, thereby isolating the stellar component of light.

The utility of this technique is demonstrated for LMC X--3, a system
which exhibits large-amplitude, but simple, variability in its
perennial thermal state.  We demonstrate that this method improves the
observability of the stellar ellipsoidal light curves and leads to an
improved determination of the orbital period.  Furthermore, the method
allows one to disentangle the OIR component of disk emission from the
reprocessed X-ray emission.  The time lags in the system are
independent of wavelength, which indicates that they originate from a
hot ring at a single, fixed radius in the disk.  We identify the
radius of this ring as the circularization radius where the tidal
stream from LMC X--3's B-star companion meets the disk, inducing a
bright shock.

By interpreting our results through the lens of $\alpha$-disk models,
we estimate the viscosity in the outer disk: Based upon the average
properties of the disk, we estimate $\alpha \sim 0.5$ to order
magnitude.  Furthermore, we unexpectedly find that $\alpha$ diminishes
as luminosity increases ($\alpha \propto \mdot^{-1/2}$).  We speculate
that this result may be related to changes in $P_{\rm gas}/P_{\rm
  mag}$ related to evolution in the mass accretion rate in the outer
disk (e.g, \citealt{Bai_Stone_2011}).

\acknowledgments

It is a pleasure to thank Poshak Gandhi, Chris Done, Ramesh Narayan,
Andy Fabian, and Joey Neilsen for helpful discussions which have
improved this work, as well as the anonymous referee, for her/his
helpful review.  We thank Catherine Brocksopp for making her B01 data
set available to us, and Xuening Bai for his input on disk theory.
Support for JFS has been provided by NASA Hubble Fellowship grant
HST-HF-51315.01.

\appendix

\section{Correlations Within the Fit}\label{append:correlations}

\begin{figure*}
{\includegraphics[clip=true, angle=90,width=15.85cm]{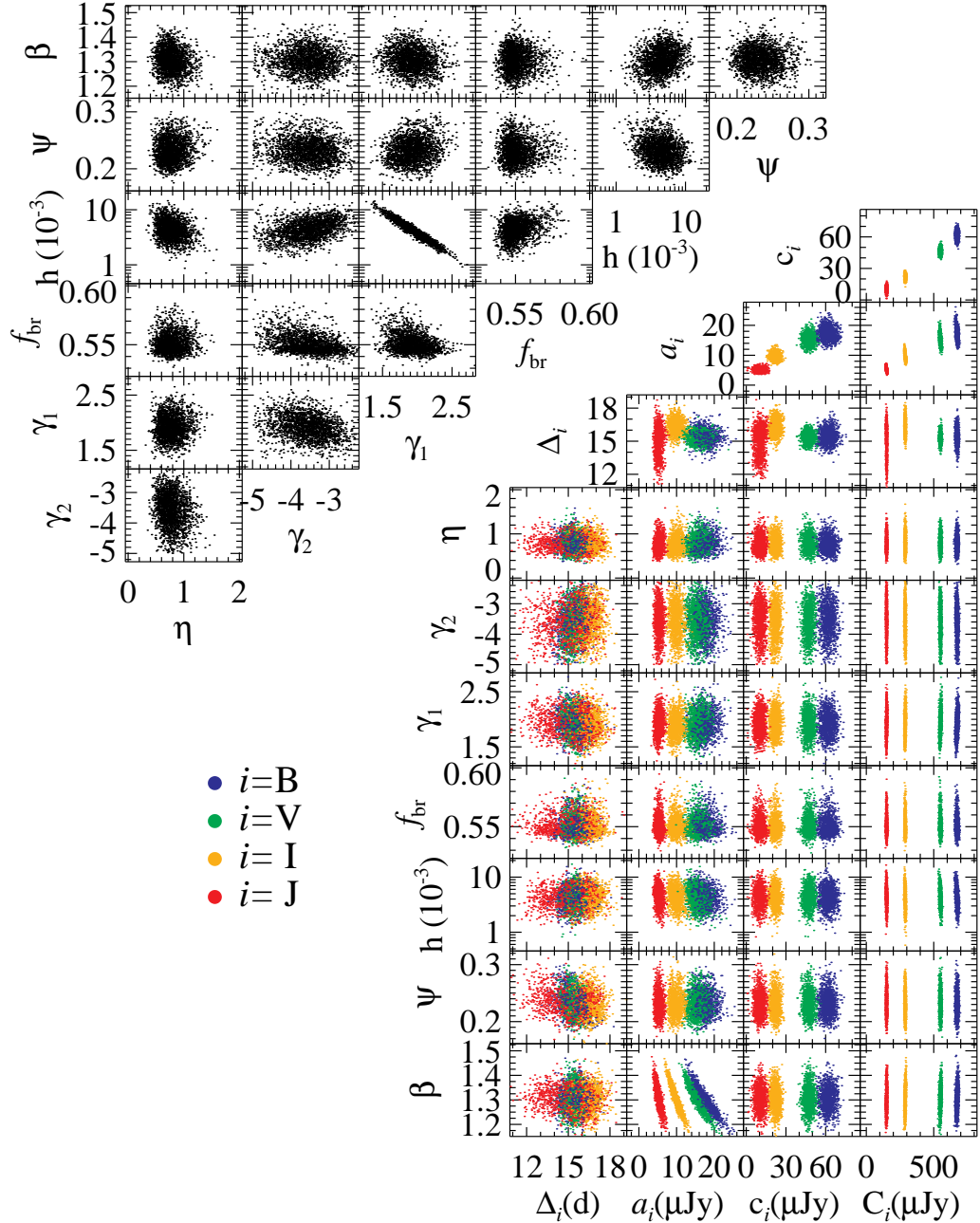}}
\caption{Correlations among selected fit parameters.  To avoid
  saturating the figure, we have plotted a randomly drawn subsample of
  the MCMC run.  Parameters which are independent for each OIR band 
  are illustrated together in color. }\label{fig:pairs}
\end{figure*}

Our model has many free parameters ($\approx 30$), and it is therefore
nontrivial to assess parameter degeneracies in the model.  In
Figure~\ref{fig:pairs} we present pairwise correlations from a
randomly-selected subsample of the MCMC chains for the principal fit
parameters of the model.  Parameters which are independently
determined for the various OIR bands are shown together in color.

Few trends are strongly evident.  The strongest is an anticorrelation
between $\gamma_1$, the index which relates the degree of X-ray
heating to luminosity, and the normalization to that scaling, $h$
(Eq.~\ref{eq:xheat}).  Also evident is a much weaker and positive correlation
between $\gamma_2$ and $h$.

Next-most prominent is an anticorrelation between $\beta$ and the disk
emission strength $a_i$, where the effect is strongest for shorter
wavelengths.  While the wavelength dependence of a given pair of model
parameters does not necessarily indicate a degeneracy in the model, it
can nevertheless be revealing.  In particular, strongly positive
trends with wavelength are evident and expected amongst parameters
that characterize the disk brightness and stellar brightness (i.e.,
panels with pairs of $a_i$, $c_i$, and $C_i$).  Conversely, the {\em
  wavelength independence} of the accretion lag, $\Delta_i$, an
important result from this work, is evident in Figure~\ref{fig:pairs}.

\section{Considering Non-Stationary Behavior}\label{section:model2}

One drawback to our approach is that although we model dynamic
evolution of the accretion-disk system, we have adopted a structure rooted in
a steady-state formalism.  We now consider an improved treatment of
non-stationary behavior with our model.  Descriptively, we better allow for
compression and expansion of the individual annuli by introducing a
term that allows for stretching action by adjacent rings.
Specifically, we modify Equation~\ref{eq:f_lag} by allowing X-ray
heating to depend on the time derivative of the inner-disk accretion
rate, $\ddot{m}_X$.  With this modification, Equation~\ref{eq:f_lag}
becomes
\begin{equation}
X_i(t) \equiv a_{i} \times \left[\frac{\mdot(t+\delta t(t)) }{ 10^{18}g\;s^{-1}}  +a_2\;\frac{\ddot{m}_X(t+\delta t(t))}{10^{18}g\;s^{-2}}\times\left(\frac{\mdot(t+\delta t(t))}{10^{18}g\;s^{-1}}\right)^{a_3}\right]^\beta,
\label{eq:mddot}
\end{equation}
where $a_2$ and $a_3$ are additional free parameters used to determine
radial expansion or compression of the flow.  This alternate
formulation has only a minor affect on the fits to the OIR light
curves; the fitting results are given in Table~\ref{tab:model2}.  The
quality of the fit is only moderately improved, $\Delta \chi^2 = 29$
for two additional free parameters.  The principal change is that the
(average) time lag inferred is smaller by $\approx 20\%$ compared to
the principal model, which is perhaps a consequence of a phase shift
induced when mixing the $\mdot$-light curve with its time derivative.
Detailed consideration of non-stationary disk modeling within the
paradigm of a bifurcated, two-state accretion flow, as applied to LMC
X--3, can be found in \citet{Cambier_2013}.

  \begin{deluxetable}{ccccccccccccccc} 
  \tabletypesize{\scriptsize} 
  \tablecolumns{      7}
  \tablewidth{0pc}  
  \tablecaption{Best Fit with Model~2}
  \tablehead{\colhead{Param.} & \colhead{Global} & \colhead{B} & \colhead{V} & \colhead{I} & \colhead{J} & \colhead{Prior}}
  \startdata  

$P$ (d)                                          &      $           1.704805\pm3\times10^{-6} $  &  \nodata & \nodata & \nodata & \nodata & Flat \\
$\beta$                                         &       $                                1.29 \pm 0.08 $  & \nodata & \nodata & \nodata & \nodata & $N(1,0.5)$ \\
$\dot{m}_{\rm break}/{\rm max}(\mdot)$   &          $                    0.546^{+ 0.012}_{- 0.006} $  & \nodata & \nodata & \nodata & \nodata & Log \\
${\rm log}_{10}(a_2)$             &         $                          0.1^{+ 0.2}_{- 0.3} $  & \nodata & \nodata & \nodata & \nodata & Flat on $a_2$ \\
 $a_3$             &                              $                                  0.4 \pm 0.4 $  & \nodata & \nodata & \nodata & \nodata & Flat \\
$\psi$                                           &      $                       0.31^{+ 0.04}_{- 0.05} $  & \nodata & \nodata & \nodata & \nodata & Flat \\
$a_X (10^{18}~g\;s^{-1})$                           &  $                                0.10 \pm 0.09 $  & \nodata & \nodata & \nodata & \nodata & Flat \\
$t_0-T_{0,{\rm Cowley}}$ (d)                        &   $                              0.001 \pm 0.014 $  & \nodata & \nodata & \nodata & \nodata & $N(0,0.01)$ \\
 $\Delta_{i}$ (d)                             &        \nodata &  $                                 12.6 \pm 1.0 $  & $                                 12.6 \pm 0.8 $  & $                                 15.4 \pm 1.2 $  & $                                 11.2 \pm 1.5 $  & Flat \\
$a_{i}  (\uJy) $                                   &   \nodata &  $                            17^{+ 4}_{- 3} $  & $                                   16 \pm 3 $  & $                                  9.9 \pm 1.9 $  & $                                  5.2 \pm 1.1 $  & Log \\
$c_{i}  (\uJy)$                                    &   \nodata &  $                                   61 \pm 6 $  & $                                   47 \pm 4 $  & $                                   22 \pm 4 $  & $                                   10 \pm 4 $  & Log \\
$C_{i}  (\uJy)$                                    &   \nodata &  $                                 677 \pm 10 $  & $                           551.^{+ 6}_{- 9} $  & $                                  287 \pm 6 $  & $                                  148 \pm 4 $  & Log \\
$\Delta m_{\rm B}$       &           $                              0.293 \pm 0.012 $  & \nodata & \nodata & \nodata & \nodata & Flat \\
$\Delta m_{\rm V}$       &           $                              0.215 \pm 0.010 $  & \nodata & \nodata & \nodata & \nodata & Flat \\
${\rm log}_{10}(h)$                                    &   $                         -2.6^{+ 0.3}_{- 0.4} $  & \nodata & \nodata & \nodata & \nodata & Log on $h$\\
$\gamma_1$                                           & $                                  2.3 \pm 0.4 $  & \nodata & \nodata & \nodata & \nodata & $N(1,0.5)$ \\
$\gamma_2$                                           & $                                 -3.5 \pm 0.9 $  & \nodata & \nodata & \nodata & \nodata & Flat \\
 $\eta$                                               &      $                                  0.8 \pm 0.3 $  & \nodata & \nodata & \nodata & \nodata & Flat \\
 ${\rm log}_{10}(\epsilon)$                       &    $                         -1.1^{+ 0.3}_{- 0.7} $  & \nodata & \nodata & \nodata & \nodata & Flat  on $\epsilon$ \\

\hline
$\rchinu$     &   2395 / 2892 \\  


\enddata
\tablecomments{Best fit model and associated 90\% credible intervals.
All notes from Table~\ref{tab:fits} are also applicable here.}
\label{tab:model2}
\end{deluxetable}


\newcounter{BIBcounter}        
\refstepcounter{BIBcounter}


\begin{thebibliography}{51}
\expandafter\ifx\csname natexlab\endcsname\relax\def\natexlab#1{#1}\fi

\bibitem[{{Arnaud}(1996)}]{XSPEC}
{Arnaud}, K.~A. 1996, in Astronomical Society of the Pacific Conference Series,
  Vol. 101, Astronomical Data Analysis Software and Systems V, ed. G.~H.
  {Jacoby} \& J.~{Barnes}, 17

\bibitem[{{Bai} \& {Stone}(2011)}]{Bai_Stone_2011}
{Bai}, X.-N., \& {Stone}, J.~M. 2011, \apj, 736, 144

\bibitem[{{Bloom} {et~al.}(2004){Bloom}, {van Dokkum}, {Bailyn}, {Buxton},
  {Kulkarni}, \& {Schmidt}}]{Bloom_2004}
{Bloom}, J.~S., {van Dokkum}, P.~G., {Bailyn}, C.~D., {Buxton}, M.~M.,
  {Kulkarni}, S.~R., \& {Schmidt}, B.~P. 2004, \aj, 127, 252

\bibitem[{{Brocksopp} {et~al.}(2001){Brocksopp}, {Groot}, \&
  {Wilms}}]{Brocksopp_2001}
{Brocksopp}, C., {Groot}, P.~J., \& {Wilms}, J. 2001, \mnras, 328, 139

\bibitem[{{Calvelo} {et~al.}(2009){Calvelo}, {Vrtilek}, {Steeghs}, {Torres},
  {Neilsen}, {Filippenko}, \& {Gonz{\'a}lez Hern{\'a}ndez}}]{Calvelo_2009}
{Calvelo}, D.~E., {Vrtilek}, S.~D., {Steeghs}, D., {Torres}, M.~A.~P.,
  {Neilsen}, J., {Filippenko}, A.~V., \& {Gonz{\'a}lez Hern{\'a}ndez}, J.~I.
  2009, \mnras, 399, 539

\bibitem[{{Cambier} \& {Smith}(2013)}]{Cambier_2013}
{Cambier}, H.~J., \& {Smith}, D.~M. 2013, \apj, 767, 46

\bibitem[{{Chiang} \& {Goldreich}(1997)}]{Chiang_Goldreich_1997}
{Chiang}, E.~I., \& {Goldreich}, P. 1997, \apj, 490, 368

\bibitem[{{Cleveland}(1979)}]{LOWESS}
{Cleveland}, W.~S. 1979, JASA, 74, 829

\bibitem[{{Cowley} {et~al.}(1983){Cowley}, {Crampton}, {Hutchings},
  {Remillard}, \& {Penfold}}]{Cowley_1983}
{Cowley}, A.~P., {Crampton}, D., {Hutchings}, J.~B., {Remillard}, R., \&
  {Penfold}, J.~E. 1983, \apj, 272, 118

\bibitem[{{Cowley} {et~al.}(1991){Cowley}, {Schmidtke}, {Ebisawa}, {Makino},
  {Remillard}, {Crampton}, {Hutchings}, {Kitamoto}, \& {Treves}}]{Cowley_1991}
{Cowley}, A.~P., {et~al.} 1991, \apj, 381, 526

\bibitem[{{Cunningham}(1976)}]{Cunningham_1976}
{Cunningham}, C. 1976, \apj, 208, 534

\bibitem[{{Durant} {et~al.}(2011){Durant}, {Shahbaz}, {Gandhi}, {Cornelisse},
  {Mu{\~n}oz-Darias}, {Casares}, {Dhillon}, {Marsh}, {Spruit}, {O'Brien},
  {Steeghs}, \& {Hynes}}]{Durant_2011}
{Durant}, M., {et~al.} 2011, \mnras, 410, 2329

\bibitem[{{Edelson} \& {Krolik}(1988)}]{EK88}
{Edelson}, R.~A., \& {Krolik}, J.~H. 1988, \apj, 333, 646

\bibitem[{{Foreman-Mackey} {et~al.}(2013){Foreman-Mackey}, {Hogg}, {Lang}, \&
  {Goodman}}]{emcee}
{Foreman-Mackey}, D., {Hogg}, D.~W., {Lang}, D., \& {Goodman}, J. 2013, \pasp,
  125, 306

\bibitem[{{Frank} {et~al.}(2002){Frank}, {King}, \& {Raine}}]{FKR}
{Frank}, J., {King}, A., \& {Raine}, D.~J. 2002, Accretion Power in
  Astrophysics (Cambridge University Press)

\bibitem[{{Fukue}(1992)}]{Fukue_1992}
{Fukue}, J. 1992, \pasj, 44, 663

\bibitem[{{Gandhi} {et~al.}(2010){Gandhi}, {Dhillon}, {Durant}, {Fabian},
  {Kubota}, {Makishima}, {Malzac}, {Marsh}, {Miller}, {Shahbaz}, {Spruit}, \&
  {Casella}}]{Gandhi_2010}
{Gandhi}, P., {et~al.} 2010, \mnras, 407, 2166

\bibitem[{Gelman \& Rubin(1992)}]{Gelman_Rubin}
Gelman, A., \& Rubin, D. 1992, Statistical Science, 7, 457

\bibitem[{{Hartmann}(1998)}]{Hartmann_book}
{Hartmann}, L. 1998, {Accretion Processes in Star Formation} (Cambridge, UK,
  Cambridge University Press)

\bibitem[{{Johnston} {et~al.}(1979){Johnston}, {Bradt}, \&
  {Doxsey}}]{Johnston_1979}
{Johnston}, M.~D., {Bradt}, H.~V., \& {Doxsey}, R.~E. 1979, \apj, 233, 514

\bibitem[{{Jones} {et~al.}(1974){Jones}, {Chetin}, \& {Liller}}]{Jones_1974}
{Jones}, C.~A., {Chetin}, T., \& {Liller}, W. 1974, \apjl, 190, L1

\bibitem[{{King}(1998)}]{King_1998}
{King}, A.~R. 1998, \mnras, 296, L45

\bibitem[{{King} {et~al.}(2007){King}, {Pringle}, \& {Livio}}]{King_2007}
{King}, A.~R., {Pringle}, J.~E., \& {Livio}, M. 2007, \mnras, 376, 1740

\bibitem[{{Kotze} \& {Charles}(2012)}]{Kotze_2012}
{Kotze}, M.~M., \& {Charles}, P.~A. 2012, \mnras, 420, 1575

\bibitem[{{Kubota} {et~al.}(2010){Kubota}, {Done}, {Davis}, {Dotani}, {Mizuno},
  \& {Ueda}}]{Kubota_2010}
{Kubota}, A., {Done}, C., {Davis}, S.~W., {Dotani}, T., {Mizuno}, T., \&
  {Ueda}, Y. 2010, \apj, 714, 860

\bibitem[{{Kupfer} {et~al.}(2013){Kupfer}, {Groot}, {Levitan}, {Steeghs},
  {Marsh}, {Rutten}, \& {Nelemans}}]{Kupfer_2013}
{Kupfer}, T., {Groot}, P.~J., {Levitan}, D., {Steeghs}, D., {Marsh}, T.~R.,
  {Rutten}, R.~G.~M., \& {Nelemans}, G. 2013, \mnras, 432, 2048

\bibitem[{{Leong} {et~al.}(1971){Leong}, {Kellogg}, {Gursky}, {Tananbaum}, \&
  {Giacconi}}]{Leong_1971}
{Leong}, C., {Kellogg}, E., {Gursky}, H., {Tananbaum}, H., \& {Giacconi}, R.
  1971, \apjl, 170, L67

\bibitem[{{Li} {et~al.}(2005){Li}, {Zimmerman}, {Narayan}, \&
  {McClintock}}]{KERRBB}
{Li}, L.-X., {Zimmerman}, E.~R., {Narayan}, R., \& {McClintock}, J.~E. 2005,
  \apjs, 157, 335

\bibitem[{{Luketic} {et~al.}(2010){Luketic}, {Proga}, {Kallman}, {Raymond}, \&
  {Miller}}]{Luketic_2010}
{Luketic}, S., {Proga}, D., {Kallman}, T.~R., {Raymond}, J.~C., \& {Miller},
  J.~M. 2010, \apj, 719, 515

\bibitem[{{McClintock} {et~al.}(2006){McClintock}, {Shafee}, {Narayan},
  {Remillard}, {Davis}, \& {Li}}]{McClintock_2006}
{McClintock}, J.~E., {Shafee}, R., {Narayan}, R., {Remillard}, R.~A., {Davis},
  S.~W., \& {Li}, L.-X. 2006, \apj, 652, 518

\bibitem[{{Miller} {et~al.}(2006){Miller}, {Raymond}, {Homan}, {Fabian},
  {Steeghs}, {Wijnands}, {Rupen}, {Charles}, {van der Klis}, \&
  {Lewin}}]{Miller_2006_H1743}
{Miller}, J.~M., {et~al.} 2006, \apj, 646, 394

\bibitem[{{Neilsen} \& {Homan}(2012)}]{Neilsen_2012}
{Neilsen}, J., \& {Homan}, J. 2012, \apj, 750, 27

\bibitem[{{Neilsen} {et~al.}(2008){Neilsen}, {Steeghs}, \&
  {Vrtilek}}]{Neilsen_2008}
{Neilsen}, J., {Steeghs}, D., \& {Vrtilek}, S.~D. 2008, \mnras, 384, 849

\bibitem[{{Orosz} {et~al.}(2014){Orosz}, {Steiner}, {McClintock}, {Buxton},
  {Bailyn}, {Steeghs}, {Guberman}, \& {Torres}}]{Jerry_LMCX3}
{Orosz}, J., {Steiner}, J., {McClintock}, J., {Buxton}, M., {Bailyn}, C.,
  {Steeghs}, D., {Guberman}, A., \& {Torres}, M. 2014, ApJ (submitted)

\bibitem[{{Penna} {et~al.}(2013){Penna}, {S{\c a}dowski}, {Kulkarni}, \&
  {Narayan}}]{Penna_2013}
{Penna}, R.~F., {S{\c a}dowski}, A., {Kulkarni}, A.~K., \& {Narayan}, R. 2013,
  \mnras, 428, 2255

\bibitem[{{Ponti} {et~al.}(2012){Ponti}, {Fender}, {Begelman}, {Dunn},
  {Neilsen}, \& {Coriat}}]{Ponti_2012}
{Ponti}, G., {Fender}, R.~P., {Begelman}, M.~C., {Dunn}, R.~J.~H., {Neilsen},
  J., \& {Coriat}, M. 2012, \mnras, 422, L11

\bibitem[{{Scargle}(1982)}]{LombScargle}
{Scargle}, J.~D. 1982, \apj, 263, 835

\bibitem[{{Shakura} \& {Sunyaev}(1973)}]{SS73}
{Shakura}, N.~I., \& {Sunyaev}, R.~A. 1973, \aap, 24, 337

\bibitem[{{Smale} \& {Boyd}(2012)}]{Smale_2012}
{Smale}, A.~P., \& {Boyd}, P.~T. 2012, \apj, 756, 146

\bibitem[{{Song} {et~al.}(2010){Song}, {Tripp}, {Wang}, {Yao}, {Cui}, {Xue},
  {Orosz}, {Steeghs}, {Steiner}, {Torres}, \& {McClintock}}]{Song_2010}
{Song}, L., {et~al.} 2010, \aj, 140, 794

\bibitem[{{Soria} {et~al.}(2001){Soria}, {Wu}, {Page}, \&
  {Sakelliou}}]{Soria_2001}
{Soria}, R., {Wu}, K., {Page}, M.~J., \& {Sakelliou}, I. 2001, \aap, 365, L273

\bibitem[{{Steeghs}(2004)}]{Steeghs_2004}
{Steeghs}, D. 2004, Astronomische Nachrichten, 325, 185

\bibitem[{{Steiner} {et~al.}(2014){Steiner}, {McClintock}, {Orosz},
  {Remillard}, {Bailyn}, {Kolehmainen}, \& {Straub}}]{Steiner_LMCX3spin}
{Steiner}, J., {McClintock}, J., {Orosz}, J., {Remillard}, R., {Bailyn}, C.,
  {Kolehmainen}, M., \& {Straub}, O. 2014, ApJ (submitted)

\bibitem[{{Steiner} {et~al.}(2010){Steiner}, {McClintock}, {Remillard}, {Gou},
  {Yamada}, \& {Narayan}}]{Steiner_2010}
{Steiner}, J.~F., {McClintock}, J.~E., {Remillard}, R.~A., {Gou}, L., {Yamada},
  S., \& {Narayan}, R. 2010, \apjl, 718, L117

\bibitem[{{Steiner} {et~al.}(2009){Steiner}, {Narayan}, {McClintock}, \&
  {Ebisawa}}]{Steiner_simpl}
{Steiner}, J.~F., {Narayan}, R., {McClintock}, J.~E., \& {Ebisawa}, K. 2009,
  \pasp, 121, 1279

\bibitem[{{Straub} {et~al.}(2011){Straub}, {Bursa}, {S{\c a}dowski}, {Steiner},
  {Abramowicz}, {Klu{\'z}niak}, {McClintock}, {Narayan}, \&
  {Remillard}}]{Straub_2011}
{Straub}, O., {et~al.} 2011, \aap, 533, A67

\bibitem[{{Subasavage} {et~al.}(2010){Subasavage}, {Bailyn}, {Smith}, {Henry},
  {Walter}, \& {Buxton}}]{Subasavage_2010}
{Subasavage}, J.~P., {Bailyn}, C.~D., {Smith}, R.~C., {Henry}, T.~J., {Walter},
  F.~M., \& {Buxton}, M.~M. 2010, in Society of Photo-Optical Instrumentation
  Engineers (SPIE) Conference Series, Vol. 7737, Society of Photo-Optical
  Instrumentation Engineers (SPIE) Conference Series

\bibitem[{{Toor} \& {Seward}(1974)}]{Toor_Seward}
{Toor}, A., \& {Seward}, F.~D. 1974, \aj, 79, 995

\bibitem[{{Wen} {et~al.}(2006){Wen}, {Levine}, {Corbet}, \& {Bradt}}]{Wen_2006}
{Wen}, L., {Levine}, A.~M., {Corbet}, R.~H.~D., \& {Bradt}, H.~V. 2006, \apjs,
  163, 372

\bibitem[{{Wilms} {et~al.}(2000){Wilms}, {Allen}, \& {McCray}}]{TBABS}
{Wilms}, J., {Allen}, A., \& {McCray}, R. 2000, \apj, 542, 914

\bibitem[{{Wilms} {et~al.}(2001){Wilms}, {Nowak}, {Pottschmidt}, {Heindl},
  {Dove}, \& {Begelman}}]{Wilms_2001}
{Wilms}, J., {Nowak}, M.~A., {Pottschmidt}, K., {Heindl}, W.~A., {Dove}, J.~B.,
  \& {Begelman}, M.~C. 2001, \mnras, 320, 327

\end{thebibliography}
\end{document}